\documentclass{elsart}
\usepackage{epsfig}

\begin{document}

\begin{frontmatter}
\title{Dynamical geometry for  multiscale dissipative particle dynamics}


\author[QMUL]{G. De Fabritiis},
\author[UCL]{P. V. Coveney}
\address[QMUL]{Centre for Computational Science, Queen Mary, University of London,\\
  Mile End Road, London E1 4NS, UK}
\address[UCL]{Centre for Computational Science, Department of Chemistry, \\
 University College  London, 20 Gordon Street, London WC1H 0AJ}

\begin{abstract}
In this paper, we review the computational aspects of a multiscale
dissipative particle dynamics model for complex fluid simulations
based on the feature-rich geometry of the Voronoi tessellation.
The geometrical features of the model are critical since the mesh
is directly connected to the physics by the interpretation of the
Voronoi volumes of the tessellation as coarse-grained fluid
clusters.
The
Voronoi tessellation is maintained dynamically in time to model
the fluid in the Lagrangian frame of reference, including
imposition of  periodic boundary conditions. Several algorithms to
construct and maintain the periodic Voronoi tessellations are
reviewed in two and three spatial dimensions and their parallel
performance discussed. The insertion of polymers and colloidal
particles in the  fluctuating hydrodynamic solvent is described
using surface boundaries.
\end{abstract}
\end{frontmatter}

\section{Introduction}

The non-equilibrium behaviour of complex fluids continues to present a major
challenge for both theory and numerical simulation. Such fluids include
multi-phase flows, particulate and colloidal suspensions, polymers,
amphiphilic fluids, including emulsions and microemulsions, and other fluids
where Brownian motion is important. Over the last decade several strategies
have been introduced both from a microdynamical point of view and from a
continuum or macro-dynamical point of view.
This article is particularly focused on mesoscopic/particulate approaches,
but the Voronoi tessellation has been employed for similar applications to
model colloidal particles and polymers by solving conventional Navier-Stokes
like equations with moving boundary conditions (see \cite{yuan98} and
references therein).

A recent contribution to the family of bottom-up approaches is the
dissipative particle dynamics (DPD) method introduced by Hoogerbrugge and
Koelman in 1992~\cite{hoog92}. Successful applications of the technique have
been made to colloidal suspensions~\cite{boek97}, polymer solutions ~\cite%
{schl95} and binary immiscible fluids~\cite{coveney96}. %
%
Dissipative particle dynamics has been shown to produce the correct
macroscopic (continuum) theory; that is, for a one-component DPD fluid, the
Navier-Stokes equations emerge in the large scale limit, and the fluid
viscosity can be computed~\cite{espanol95,marsh97}. However, even though
dissipative particles have generally been viewed as clusters of molecules,
the cluster of molecules are not considered as a thermodynamic system
itself.  In a following development of the DPD model, Flekk{\o }y and
Coveney~\cite{flekkoy99,flekkoy00,defabritiis01} and Serrano and Espa{\~n}ol
\cite{serrano01} derived a new model which provides a precise definition of
the term `mesoscale' and enables us to retrieve numerically the specific
thermodynamics of the solvent in the sense that the exact Gibbs equilibrium
is approached, the entropy is an increasing function in time, and the total
energy is exactly conserved \cite{serrano02}. %
%
In this approach, the fluid dissipative particles are defined as cells on a
Voronoi lattice with variable sizes and masses; the Voronoi tessellation
must be updated at each time step under the dynamics of the model. The
scheme enables one to select any desired local scale for the mesoscopic
description of a given problem. Indeed, the method may be used to deal with
situations in which several different length scales are simultaneously
present.

In this article, we discuss the computational aspects of the model and its
computational overhead due to the Voronoi tessellation; in particular, we
present the development of an efficient C++ code for the implementation of
the new fluid particle model. Our discussion concentrates on the challenging
aspects of maintaining the Voronoi tessellation over time, with various
kinds of boundary conditions.

\section{Multiscale dissipative particle dynamics}

In our case, we follow the model described in detail in \cite%
{serrano01,serrano02}, but we apply its isothermal version, which has the
advantage  of greater simplicity. In fact, if heat diffusion is not
important for the system under consideration, then the isothermal model
furnishes an efficient alternative, because several simplifications are
possible in the equations (for instance, the It\^{o}-Stratonovich terms all
vanish). Defining the state of a fluid particle by its position, mass and
momentum $\mathbf{x}=\{\mathbf{r}_{i},M_{i},\mathbf{P}_{i}\}$, the resulting
equations for the isothermal model are
\begin{eqnarray}
d\mathbf{r}_{i} &=&\mathbf{v}_{i}dt,  \nonumber \\
d{M}_{i} &=&\sum_{j}\frac{\rho _{i}+\rho _{j}}{2}\mathbf{C}_{ij}\!\cdot \!%
\mathbf{v}_{ij}dt,  \nonumber \\
d\mathbf{P}_{i} &=&\sum_{j}\frac{A_{ij}}{2}\left( p_{ji}\mathbf{1}+\mathbf{%
\Pi }_{i}+\Pi _{i}\mathbf{1}\right) \cdot \mathbf{e}_{ij}dt+d\widetilde{%
\mathbf{P}}_{i}  \nonumber \\
&+&\sum_{j}\frac{\rho _{i}+\rho _{j}}{2}\frac{\mathbf{v}_{i}+\mathbf{v}_{j}}{%
2}\mathbf{C}_{ij}\!\cdot \!\mathbf{v}_{ij}dt.  \label{sdeiso}
\end{eqnarray}%
\noindent

\noindent where $\mathbf{v}_{i}=\mathbf{P}_{i}/M_{i}$ is the velocity, $%
\rho_{i}=M_{i}/V_{i}$ is the mass density,
and $d$ is the spatial dimension. For a general variable $x$, we write $%
x_{ij}=x_{i}-x_{j}$. The pressure $p_{i}$ is given through the equilibrium
equations of state once the temperature $T$ is assigned. We have also
introduced geometric quantities arising from the Voronoi construction: $%
A_{ij}$ is the area (length in 2D) of the face between cells $(i,j)$, $%
\mathbf{e}_{ij}=(\mathbf{r}_{i}-\mathbf{r}_{j})/r_{ij}$ with $r_{ij}=|%
\mathbf{r}_{i}-\mathbf{r}_{j}|$ is the unit vector normal to the face $(i,j)$
and, finally, $\mathbf{C}_{ij}=\frac{A_{ij}}{r_{ij}}\left( \mathbf{A}_{cm}-%
\frac{\mathbf{r}_{i}+\mathbf{r}_{j}}{2}\right) $ is a geometrical vector
parallel to the face $(i,j)$. The dissipative stress tensor is
\begin{equation}
\mathbf{\Pi}_{i}^{\alpha\beta} =\frac{\eta_{i}}{V_{i}}\left[ \frac{1}{2}%
\sum_{j}A_{ij}[\mathbf{e}_{ij}^{\alpha}\mathbf{v}_{j}^{\beta}+\mathbf{e}%
_{ij}^{\beta}\mathbf{v}_{j}^{\alpha}]-\frac{1}{D}\delta
^{\alpha\beta}\sum_{j}A_{ij}\mathbf{e}_{ij}\!\cdot\!\mathbf{v}_{j}\right] ,
\end{equation}
while the fluctuations in momentum are 
\begin{equation}
d\widetilde{\mathbf{P}}_{i} =\sum_{j}\frac{1}{2}A_{ij}\mathbf{d\widetilde {%
\sigma}}_{j}\cdot\mathbf{e}_{ij},
\end{equation}
where the stress tensor is 
\begin{equation}
\mathbf{d\widetilde{\sigma}}_{i} =\left( 4k_{B}T_{i}\frac{\eta_{i}}{V_{i}}%
\right) ^{\frac{1}{2}}d\overline{\mathbf{W}}_{i}^{S}+\left( 2Dk_{B}T_{i}%
\frac{\xi_{i}}{V_{i}}\right) ^{\frac{1}{2}}tr[d\mathbf{W}_{i}].
\end{equation}
\noindent Here, $d\mathbf{W}_{i}$ is a matrix $(d\times d)$ of independent
Wiener increments, i.e. $dW_{i}^{(\alpha,\beta)}$ has zero mean and variance
$dt$ and is uncorrelated to any other random variables, $d\overline {\mathbf{%
W}}_{i}^{S}=\frac{1}{2}\left( d\mathbf{W}_{i}+d\mathbf{W}_{i}^{t}\right) -%
\frac{1}{d}tr\left[ d\mathbf{W}_{i}\right] \mathbf{1}$ is a symmetric
traceless matrix. 


The SDEs (\ref{sdeiso}) can be integrated numerically using any stochastic
integrator, such as Euler, Heun or higher order schemes \cite{kloeden92}.
For our simulations, we used an integrator based on a generalisation of the
Trotter formula to the stochastic case \cite{defatrotter} which proved to be
simple to apply and very efficient.

\section{Voronoi tessellation}

\begin{figure}[tb]
\begin{center}
\centerline{\mbox{\epsfig{file=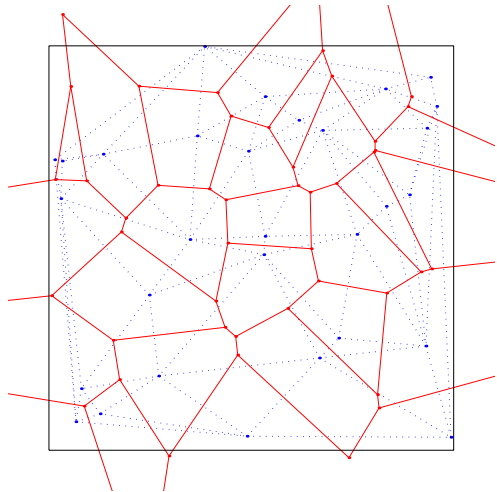,height=8cm}}}
\end{center}
\caption{The tessellation for an unbounded domain. The continuous lines are
the Voronoi edges, while the points represent the dissipative particles or
Delaunay vertices. The dotted lines are the edges of the Delaunay
triangulation and the missing Voronoi edges are infinite edges. The square
is the simulation domain.}
\label{fluidpfig}
\end{figure}

The model is centrally based on the Voronoi tessellation. The
Voronoi tessellation is simply defined by a set of points
$P=\{p_{1},...,p_{N}\}$ in Euclidean space; a partition of the
space assigning every point to its nearest site is called a
Voronoi tessellation. The Voronoi cell $V(p_{i})$ consists of all
the points at least as close to $p_{i}$ as to any other site
\[
V(p_{i})=\{x:\left| p_{i}-x\right| \leq \left| p_{j}-x\right| ,\forall
\;j\neq i\}.
\]
\noindent The Voronoi tessellation can be computed through the Delaunay
triangulation, which is the triangulation that maximises the minimum
internal angle among the triangles \cite{rourke}. In terms of the fluid
dissipative particles, the Delaunay triangulation indicates the
neighbourhood of each fluid particle in the tessellation and the Voronoi
tessellation indicates its spatial extension (see Figure \ref{fluidpfig}).
By first triangulating the set of points  $P$ with the Delaunay
triangulation, the Voronoi tessellation can then be constructed by using the
Delaunay-Voronoi duality~\cite{rourke}, i.e. to each Delaunay vertex
corresponds a Voronoi region and to each Delaunay edge corresponds a Voronoi
boundary surface or line in two dimensions.


The physical model requires maintenance of the Voronoi tessellation for each
time step following the dynamics in the Lagrangian frame of reference and
specification of the boundary conditions for the tessellation. In our
opinion, direct coding of the construction of the Voronoi tessellation is
not desirable. This still is in its own right a research field in
computational geometry in order to obtain a stable and efficient
triangulation. It involves writing exact number types to check the validity
of predicates on points (for instance the insphere property for four points)
and combinatorial aspects to handle the data structure. Nevertheless,
writing an inefficient and unstable code to construct the Voronoi
tessellation is quite simple, but the result will likely be orders of
magnitude slower and unstable in most practical situations. Whilst there are
several public libraries \cite{triangle,cgal} that are able to compute the
Voronoi-Delaunay tessellation for a given set of points. However, the
algorithmic complexity of the construction is at best $N\log N$, where $N$
is the number of points, and thus it is crucial to use an efficient library
to match that complexity. We were unable to find any library able to
dynamically change the tessellation and impose periodic boundary conditions.

Previously, in our research group, the \textit{Triangle} library~\cite%
{triangle} was employed for the construction of the Voronoi tessellation~%
\cite{fabritiis99}. Although the \textit{Triangle} library does not support
three dimensional Voronoi tessellations, it is the fastest library known to
us in two dimensions. \emph{Triangle} is based on the \textit{%
divide-and-conquer} algorithm \cite{rourke}, which allows one to compute the
tessellation given the complete list of points. It furnishes as well an
incremental algorithm, which provides the functions to construct the
tessellation point by point.

With \emph{Triangle} the Voronoi tessellation was maintained over time by
deleting the set of points $P$ from the tessellation, determining the new
positions of the points and then computing the tessellation for the new
points, for every time-step. Given the algorithmic complexity of the
construction of Voronoi tessellations, this is clearly not an efficient way
to maintain the tessellation over time; it is desirable to be able to
locally \textit{restore} the tessellation at each time-step, thus avoiding
deleting and then re-computing the entire tessellation.

Subsequent awareness of the Computational Geometry Algorithms Library (CGAL)
\cite{cgal} gave us a homogeneous environment for two and three dimensional
tessellations. CGAL incorporates a wide-range of function modifiers,
including the insertion or removal of specific points in a tessellation,
thus offering a highly flexible code for the construction and maintenance of
a Voronoi tessellation over time. Consequently, we are currently employing
the CGAL library for the development of the Voronoi DPD code in two and
three spatial dimensions.

\section{Boundary conditions}

In this section, we describe possible implementations of periodic
and wall boundary conditions in two and three dimensions based on
the insertion-removal of image cells (points) around the
simulation box. The insertion of polymers and colloidal particles
is described using boundary surfaces. We present these algorithms
in two dimensions because our three dimensional implementation is
a straightforward application of the same techniques. The library
takes care of the extra complexity of handling the three
dimensional tessellation. Differences  are pointed out in Section
\ref{effimages}.

\subsection{Periodic boundary conditions}

\begin{figure}[tbp]
\begin{center}
\centerline{\mbox{\epsfig{file=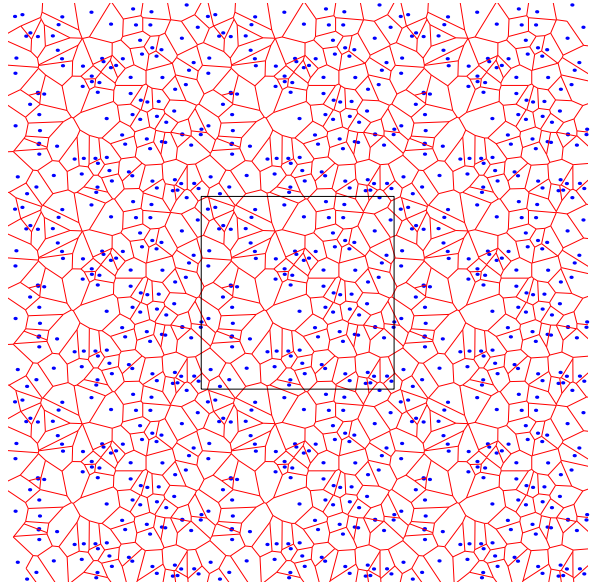,height=8cm}}}
\end{center}
\caption{The simplest way to construct a periodic tessellation. The points
in the simulation region are copied to the corresponding positions in the
eight surrounding boundary regions to produce a periodic Voronoi
tessellation.}
\label{sim9fig}
\end{figure}

Boundary conditions can be implemented by inserting `image points' around
the boundaries of the space domain. We call image points the points of the
tessellation which lie outside the simulation box and represent images of
centres of fluid particles inside the simulation box. The simplest way to
achieve a periodic tessellation is to copy the simulation region to the
eight surrounding boundary regions (see Figure \ref{sim9fig}).

%
However, this requires that the tessellation has to be computed for a total
of $9N$ points if $N$ is the number of points to be triangulated in the
simulation region. It is obviously not necessary to triangulate this number
of image points to maintain the periodic boundary conditions and in light of
the algorithmic complexity of the construction of the tessellation, it is
crucial to minimise the number of image points. Clearly, it is only the
points that lie `near' a boundary face that \textit{must} be copied to the
boundary regions. To isolate these `boundary points' we use the following
geometrical criterion: \textit{a point is in the neighbourhood of the
boundary if the presence of another point outside the domain can affect its
Voronoi region.} This criterion can be implemented with a simple algorithm.
This is given in Algorithm 1.
\begin{table}[tb]
\begin{description}
\item[Algorithm 1 (PERIODIC BOUNDARY CONDITIONS)]
\end{description}
\begin{enumerate}
\vspace{0.5pt}
 \sf{ \footnotesize{
\item Insert the corner points of the simulation domain
\item Triangulate (Move) $N$ points
\item {\bf for} every finite Delaunay triangle $(p_{1},p_{2},p_{3})$
\begin{description}
\item \{
\begin{description}
\item Compute the circumcircle $C(V)$
\item {\bf if} $C(v)$ intersects a boundary face(s)
\item \{
\item ~~~~~~Copy $p_{1}, p_{2}$ and $p_{3}$ to the boundary region(s)
\item ~~~~~~of the opposite face(s)
\item \}
\end{description}
\item \}
\end{description}
\item Remove the corner points
} }
\end{enumerate}
\end{table}

In the first step of the algorithm, $N$ points are triangulated in the
simulation region. In step $2$, the following geometrical properties of the
Delaunay-Voronoi tessellation \cite{rourke} are employed:

\begin{itemize}
\item Each node $p_{i}$ of the Delaunay triangulation corresponds to a
Voronoi region $V(p_{i})$.

\item If $v$ is a Voronoi vertex at the junction of $%
V(p_{1}),V(p_{2}),V(p_{3})$ then $v$ is the centre of the circle $C(v)$
determined by the points $p_{1},p_{2}$ and $p_{3}$.

\item $C(v)$ is the circumcircle for the Delaunay triangle $%
(p_{1},p_{2},p_{3})$ containing no other points $p_{i}$.
\end{itemize}

\noindent It follows from the Delaunay-Voronoi properties above that for a
Delaunay triangle $(p_{1},p_{2},p_{3})$, if the circumcircle $C(v)$ does
\textit{not} intersect a boundary face of the domain, then no image point
can affect the Voronoi regions $V(p_{1})$, $V(p_{2})$ and $V(p_{3})$ and
hence, according to our geometrical criterion, $p_{1}, p_{2}$ and $p_{3}$ do
not have to be copied to the boundary regions. However, if $C(v)$ \textit{%
does} intersect a boundary face, then an image point \textit{can} affect the
Voronoi regions $V(p_{1})$, $V(p_{2})$ and $V(p_{3})$. Unfortunately, it is
not possible to know if any boundary points will produce an image point
within $C(v)$ until the tessellation for these boundary points is
constructed. Nevertheless, a good upper-bound can be given on the number of
points that must be copied by considering the worst situation that for any
Delaunay triangle $(p_{1},p_{2},p_{3})$ whose circumcircle $C(v)$ interests
a boundary face, an image point will always be located within $C(v)$. In
this case $p_{1}, p_{2}$ and $p_{3}$ must be copied to the boundary region
of the opposite face; if $C(v)$ intersects the corner (i.e. two boundary
faces), $p_{1}, p_{2}$ and $p_{3}$ must be copied to the boundary regions of
the two opposite faces and also to the boundary region of the opposite
corner (see Figure \ref{sim9fig}).


The points which belong to the convex hull of the triangulation have their
Voronoi centre $v$ positioned at infinity and thus it is not possible to
compute $C(v)$ in these cases. The solution we applied is to insert the
corner points of the periodic boundary regions in order to create an
external convex hull. After the image points are calculated the corners
points are removed. This is not needed when the periodic boundary condition
is already set and we want to maintain it, because the image points will
form the convex hull.

\begin{figure}[tbp]
\begin{center}
\vspace*{.05in} \centerline{\mbox{\psfig{file=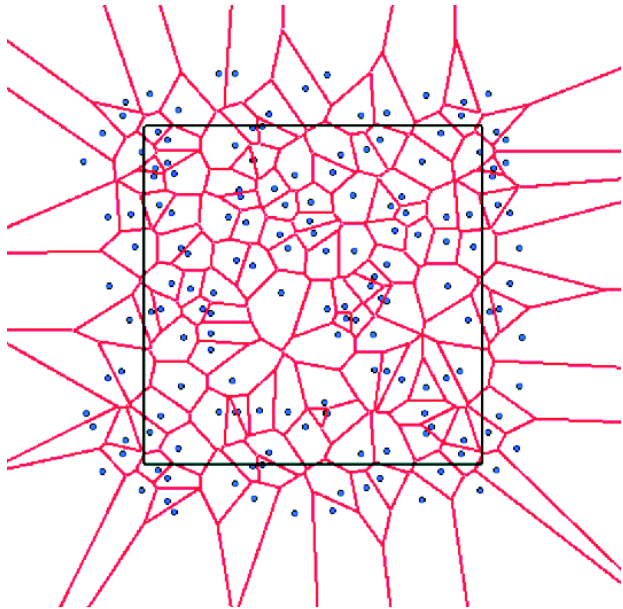,height=8cm}}}
\end{center}
\caption{Implementation of periodic boundary conditions using Algorithm 1.}
\label{periodic1ps}
\end{figure}

The periodic boundary tessellation produced by Algorithm 1 is shown in
Figure \ref{periodic1ps}. Thus Algorithm 1 avoids copying all the points
and, in assuming the worst case scenario that an image point is always
located within $C(v)$ for any boundary point, this algorithm reduces the
number of image points to the minimum upper-bound possible.

\subsection{Fixed (solid) wall boundary conditions}

\label{wallsect}

\begin{figure}[tb]
\begin{center}
\vspace*{.05in} \centerline{\mbox{\psfig{file=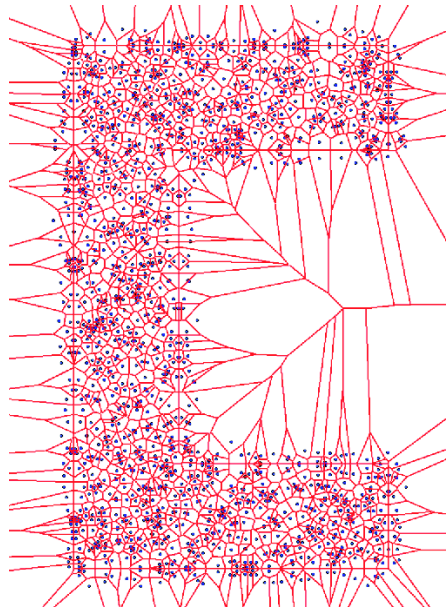,height=8cm}}}
\end{center}
\caption{Implementation of fixed (solid) wall boundary conditions using
Algorithm 2.}
\label{fixedwall}
\end{figure}

A possible way to implement fixed (solid) wall boundary conditions for the
Voronoi tessellation is to introduce image points into the boundary regions
which correspond to the reflection of the boundary points in their boundary
face(s) (see Figure \ref{fixedwall}). The interaction between a boundary
particle and the `wall' is then handled by the interaction between the
particle and its `mirror-image' particle. We have successfully implemented
fixed wall boundary conditions for the Voronoi tessellation with Algorithm 2.

\begin{table}[tb]
\begin{description}
\item[Algorithm 2 (FIXED WALL BOUNDARY CONDITIONS)]
\end{description}
\begin{enumerate}
\vspace{0.5pt}
 \sf{ \footnotesize{
\item Insert the corner points of the simulation domain
\item Triangulate $N$ points.
\item {\bf for} every finite Delaunay triangle $(p_{1},p_{2},p_{3})$
\begin{description}
\item \{
\begin{description}
\item Compute the circumcircle $C(V)$
\item {\bf if} [$C(v)$ intersects a boundary face AND $(p_{1},p_{2},p_{3})$ lies on
the correct side of the boundary]
\item \{
\item ~~~~~{\bf for} each Delaunay vertex, {\bf if} the perpendicular projection of
the vertex on to the supporting line is also on the boundary face
segment, reflect the vertex in the boundary face
\item \}
\end{description}
\item \}
\end{description}
\item Remove the corner points
\label{periodic1} } }
\end{enumerate}
\end{table}

The algorithm follows the same principle as Algorithm 1; we first insert the
corner points of the simulation domain, and then triangulate our $N$ points,
therefore enabling us to compute the circumcircle of every Delaunay triangle
in the simulation domain. Next in step 3, we isolate all of the boundary
points. However, the criterion for a boundary point for fixed wall boundary
conditions is more complicated than that for periodic boundary conditions.
For any circumcircle $C(V)$ which intersects a boundary face, two additional
requirements must be satisfied before $p_{1}, p_{2}$ or $p_{3}$ can be
isolated as boundary points.


Firstly, for a given Delaunay triangle $(p_{1},p_{2},p_{3})$, if $C(V)$
intersects a boundary face, it is possible for $(p_{1},p_{2},p_{3})$ to lie
on the wrong side of the boundary. For example, consider the fixed wall
boundary conditions for the `C' shaped domain in Figure \ref{fixedwall}.
Suppose we are isolating the particles that lie in the neighbourhood of the
first `shorter' horizontal boundary face from the bottom in Figure \ref%
{fixedwall}. It is possible for the circumcircle of a Delaunay triangle $%
(p_{1},p_{2},p_{3})$ lying \textit{above} this boundary face to intersect
this boundary and yet clearly $(p_{1},p_{2},p_{3})$ cannot physically lie in
the neighbourhood of this boundary. Thus, if $C(V)$ intersects a boundary
face, we must then check that $p_{1}, p_{2}$ and $p_{3}$ lie on the correct
side of the boundary; if we are traversing the boundary faces in a clockwise
direction then $p_{1}, p_{2}$ and $p_{3}$ are on the correct side of the
boundary face if they lie to the right-hand side of the boundary. Similarly,
an anti-clockwise traversal of the boundary faces requires that $p_{1}, p_{2}
$ and $p_{3}$ lie to the left-hand side of the boundary.

If these first two requirements are met, then we must check one final
geometrical property. We refer again to the first `shorter' horizontal
boundary face up from the bottom of the `C' shaped domain in Figure \ref%
{fixedwall}. Suppose for a Delaunay triangle $(p_{1},p_{2},p_{3})$, the
circumcircle $C(V)$ intersects this boundary face \textit{and} $p_{1}, p_{2}$
and $p_{3}$ lie on the correct side of the boundary. Consider the supporting
line of this boundary face (that is, the infinite line which is parallel to
the boundary face segment and passes through it) and the perpendicular
projection of the Delaunay vertices $p_{1}, p_{2}$ and $p_{3}$ onto this
supporting line. For the majority of cases, these three projections will
also lie on the boundary face segment in which case $p_{1}, p_{2}$ and $p_{3}
$ can be reflected in the boundary face and are thus boundary points.
However, for Delaunay triangles whose circumcircle intersects the left-end
of this boundary face, it is possible that some or even all of these
projections may not lie on the boundary face segment, that is, they do not
have a reflection in this boundary face and hence are not boundary points.
Thus, if $C(V)$ intersects a boundary face and $p_{1}, p_{2}$ and $p_{3}$
lie on the correct side of the boundary, we must then determine if the
perpendicular projection of each Delaunay vertex onto the supporting line is
also on the boundary face segment. For those that are, they are boundary
points.

Having isolated the boundary points with the above criterion, we then insert
the reflection of these points in their boundary face(s). We note that if $%
C(V)$ intersects a corner (i.e. two boundary faces) then the two additional
geometrical requirements detailed above must be applied separately to each
boundary. Finally in step 4 of Algorithm 2, we remove the corner points from
the simulation.

This algorithm can be applied to any two dimensional polygon that can be
described by a set of vectors. We show in Figure \ref{fixedwall} the fixed
wall boundary tessellation for a `C' shaped domain, obtained using Algorithm
2. We note that for corners where the angle between the two boundary faces
\textit{in the fluid} is greater than $180^0$, the actual corner is not
produced exactly. This is because near the corner, the insertion of an image
point above one of the boundary faces will affect the Voronoi region of the
other boundary face (see Figure \ref{fixedwall}). However, for realistic
simulations involving tens of thousands of points, the `approximate' corner
obtained is perfectly sufficient.

\subsection{Boundary surfaces}

One final aim of this research is to add specificity to the model by
inserting polymers and colloidal particles in the fluctuating hydrodynamic
solvent based on the Voronoi dissipative particle dynamics model. Then, the
rheological properties of the resulting system can be computed via direct
simulations of shear flows.

\begin{figure}[p!]
\vspace*{.05in} \centerline{\mbox{\psfig{file=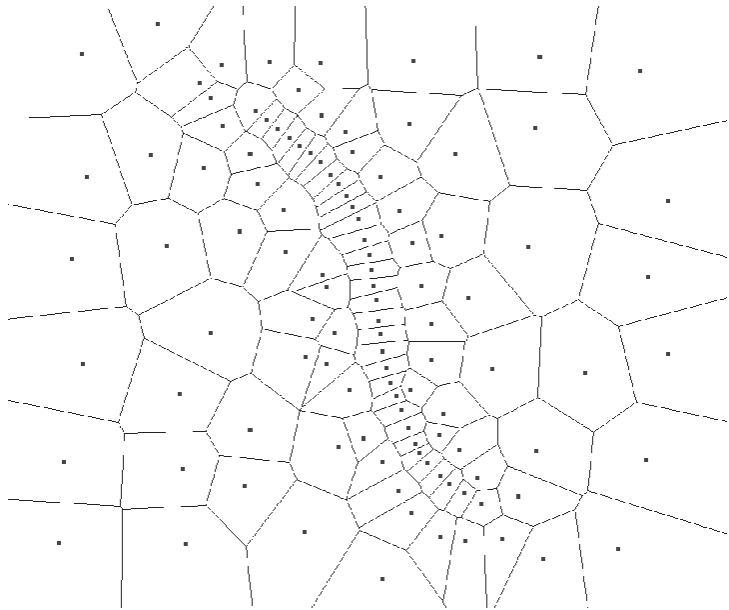,height=8cm}}}
\caption{ Structure of a polymeric fluid according to the multiscale
dissipative particle dynamics method. The resolved length scale of the
polymer is much finer than that of the fluid. In the fluid itself different
length scales are present to handle polymer-fluid interactions.}
\label{polymerfig}
\end{figure}

\begin{figure}[p!]
\vspace*{.05in} \centerline{\mbox{\psfig{file=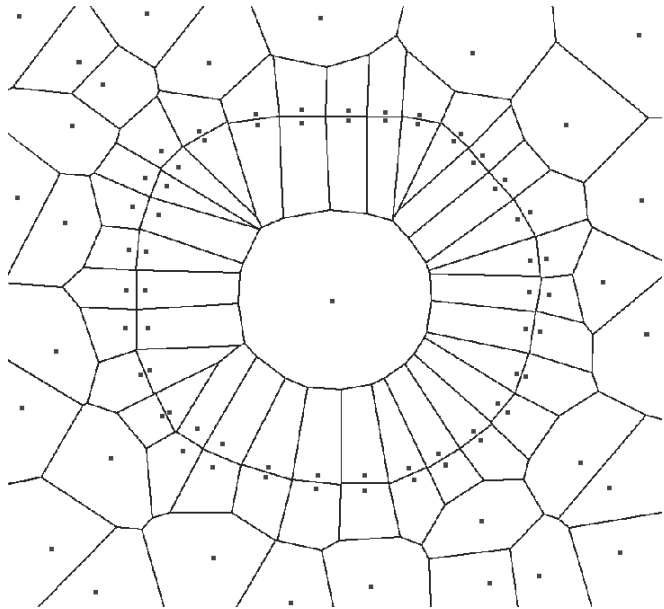,height=8cm}}}
\caption{ Structure of a colloidal particle according to the multiscale
dissipative particle dynamics method. The surface of the colloid is resolved
by inserting two points for each surface element. The central point
represents the colloid's centre of mass and is used to make the tessellation
more stable.}
\label{colloidfig}
\end{figure}

Polymers and colloidal particles are inserted in the Voronoi tessellation
using surface boundaries formed by chains of Voronoi centres. This is
represented in Figures \ref{polymerfig} and \ref{colloidfig}. The boundary
surface are fixed giving a couple of Voronoi centres sufficiently close to
form a Voronoi surface, which represents the boundary. How close they need
to be is computed from the typical scale of the system or the smallest
Voronoi fluid particle.

The physical meaning of the two Voronoi centres depends on their position.
The Voronoi cell internal to the polymer or the colloid is a coarse-grained
representation of a part of the polymer. The Voronoi centre in the fluid is
a fluid particle which does not slip on the colloidal/polymer surface.

%
%
This description of surface boundary could be used as well for wall
boundaries, with the disadvantage that an higher number of cells is needed
to construct the boundary compared to the approach presented in Section \ref%
{wallsect}, but with the advantage of an easy construction and maintenance
of the boundary during the dynamics. In fact, once the two cells forming the
boundary surface are kept at a fixed distance, then the maintenance of the
tessellation does not involve any particular extra work.

\subsection{Efficiency of image point boundary conditions}

\label{effimages}

The efficiency of the construction of periodic/solid boundary conditions
using image points depends on the number of images that we need to insert.
Intuitively, this is a limitation when we tessellate a small number of
points, because in this case the surface:volume ratio is high. We
quantitatively measured the number of images needed to impose boundary
conditions in two and three spatial dimensions.

The results are reported in Table \ref{imagepoints}. 
%
%
%
%
%
%
%
\noindent First, we note that while in two dimensions we have 8 neighbour
regions around the box, in three dimensions there are 26. The effect of this
geometrical change is evident. The number of shell points represents the
number of points that have at least one image. The number of fluid particles
which are in the core, and therefore not affected by boundary conditions, is
of course $N=N_{fp}-N_{shell}$, where $N_{fp}$ is the number of fluid
particles. Each point in the shell could be copied to more than one
neighbour region. The number of images is the number of points that are
actually inserted outside the simulation box. The number of extra images
gives the number of point images that are not connected to any point inside
the simulation domain, i.e. their presence does not affect the periodic
boundary at all. The small number of extra images compared to the number of
fluid particles shows that the criterion used to select image points is very
effective.

Obviously, it is important that the number of images, which represent a
significant computational overhead, is small compared to the number of
points in the box. This is easily obtained in two dimensions where for 1000
points we have 190 images (less than 20\%). However, in three dimensions, to
have the same percentage we need more than 50000 points. This severely
limits the performance of the current code with full periodic boundary
conditions in three dimensions.

\section{Maintaining the Voronoi tessellation dynamically}

The dynamical maintenance of the tessellation is based on the reconnection
algorithm (see \cite{reconnection}). The reconnection algorithm locally
restores the Delaunay property on a valid triangulation. It is very
effective, but we need to guarantee that after all the points are moved the
tessellation does not present invalid triangles (tetrahedra). It is not
simple to assure this condition. Usually, by reducing the time step it is
possible to control the validity of the triangulation. However, in our
opinion this method is not really stable, and brings with it the additional
problem that the reduction of the time step effectively slows down the code.
In three dimensions, we prefer to locally restore the tessellation moving
one point by one. We are thus able to check that the triangulation is still
valid and then to use asynchronous dynamics to update different fluid
particles with different time steps. This last option is particulary
interesting for a multiscale model like this. In Appendix A,
we detail the code for the maintenance of the tessellation in two and three
dimensions realised with the CGAL library \cite{cgal}.

\subsection{Local reconnection algorithm in two dimensions}

Given the algorithmic complexity of the construction of a Voronoi
tessellation for $N$ points $(N\log N)$, it is crucial that we are able to
efficiently maintain the Voronoi tessellation over time. We can make
significant improvements by utilising the flexibility of the CGAL library.
We are able to \textit{restore} the Delaunay Triangulation (and thus the
Voronoi tessellation) locally with the following algorithm (an
implementation for the CGAL library is shown in Appendix A),
which is an extension of the algorithm used by CGAL for restoring the
Delaunay triangulation when individual points are inserted or removed.

\begin{table}[tb]
\begin{description}
\item[Algorithm 3 (RESTORE DELAUNAY)]
\end{description}
\begin{enumerate}
\vspace{0.5pt}
 \sf{ \footnotesize{
\item Compute $\delta t$ such that no point can move outside its Voronoi region
\item Move the points
\item {\bf for} every finite Delaunay triangle $(p_{1},p_{2},p_{3})$
\begin{description}
\item \{
\begin{description}
\item Compute the circumcircle $C(v)$
\item {\bf if} $C(v)$ does not satisfy the empty circle property
\item \{
\begin{description}
\item ~~~Triangulate the quadrilateral formed by $p_{1}, p_{2}, p_{3}$ and the
 point lying within $C(v)$ the other way round
\end{description}
\item \}
\end{description}
\item \}
\end{description}
} }
\end{enumerate}
\end{table}

We now consider the individual steps of the algorithm. In the first step we
ensure that no point can move outside its Voronoi region by restricting the
size of the time-step.  This is achieved by iterating over the edges of the
Delaunay triangles  and computing the time-step for each edge such that
neither point can move more than a quarter the distance towards the other
one by setting $\delta t$ to the minimum of all these values. After
integrating the stochastic Langevin equations of motion, we then update the
position of each point in step $2$. It is clear that some triangles may now
be invalid Delaunay triangles and so we restore the triangulation to a
Delaunay triangulation in step $3$. We achieve this as follows: every
triangle in the simulation domain has three neighbouring triangles, that is,
three neighbouring vertices (by definition, the triangles that belong to the
convex hull have the infinite vertex as one of their three neighbouring
vertices). A triangle is a valid Delaunay triangle if it satisfies the empty
circle property, that is, if none of its neighbouring vertices lies inside
its circumcircle. Therefore, we iterate over the finite triangles in the
triangulation and check the empty circle property for every triangle; for
any triangle $(p_{1},p_{2},p_{3})$ which does not satisfy the empty circle
property due to a neighbouring vertex, say $q$, located within its
circumcircle, we simply triangulate the quadrilateral formed by $p_{1},
p_{2}, p_{3}$ and $q$ the other way round. We note that Algorithm 3 must be
implemented such that any \textit{new} triangles formed are subsequently
visited by the iteration process.

Clearly, there is far less work involved in restoring the Delaunay
triangulation than with re-computing the whole tessellation for every
time-step. Indeed, the above restore Delaunay algorithm is a very efficient
way of maintaining a two-dimensional Voronoi tessellation over time.

\subsection{Reconnection algorithm in three dimensions}

\label{rec3d}

In three dimensions, instead of performing a sequence of geometrical flips
of the facets in order to locally restore the Delaunay property, we check
the validity of the insphere property for all the facets and eventually, if
the test fails, we move back the point to the preceding position, remove it
and insert in the new position. Algorithm 4 shows this procedure.

\begin{table}[tb]
\begin{description}
\item[Algorithm 4 (RECONNECTION DELAUNAY)]
\end{description}
\begin{enumerate}
\vspace{0.5pt}
 \sf{ \footnotesize{
\item {\bf for} every point
\begin{description}
\item \{
\begin{description}
\item Move the point
\item Compute the insphere property for the facets around it
\item {\bf if}  does not satisfy the empty sphere property
\item \{
\begin{description}
\item ~~~Move back the point
\item ~~~Remove it
\item ~~~Insert the point in the new position
\item ~~~Return
\end{description}
\item \}
\end{description}
\item \}
\end{description}
} }
\end{enumerate}
\end{table}

For a standard system of fluid particles (see section \ref{setup}), the
number of topological events is usually small, around 10\%, therefore the
expensive remove-insert procedure is performed only on a very few points. To
improve performance the remove-insert task can be used only in the case of
invalid triangulation, flipping the facets to restore the Delaunay property
for the others. We note that Algorithm 4 spends a large amount of its
computational time without modifying anything in the triangulation if it
does not not adjust the position of the fluid particle. Because of this, the
algorithm is very suited for a data-parallel parallelisation. In fact, a
processor acting on a fluid particle does not invalidate the memory of any
other processor, while the few topological changes can be stored in a queue
and performed sequentially by each processor.

\subsection{Asynchronous time dynamics}

The physical model does not present any restrictions regarding its temporal
update. In principle, but not actually implemented in the code, we can add
multiscaling in time to multiscaling in space. To each fluid particle is
associated a length scale given by its volume, while its time scale can be
fixed by its velocity. Therefore, we can set the time scale for each fluid
particle and move it on this basis. Sequential temporal updating has already
been studied in particle based methods and solved with a hierarchy of time
scales of power of two \cite{hernquist89}. In practice, given the time scale
for fluid particle $i$, $dt_i$, and the largest time scale $dt_s$, then the
time steps are chosen such that
\[
dt_i=\frac{dt_s}{2^n_i},
\]
where $n_i$ refers to the time bin of particle $i$. Synchronisation is
maintained at the end of each large time step $dt_s$. Then a new large time
scale is computed, and so on. The Voronoi local maintenance algorithm can
take advantage of this time advancement given its locality.

\section{Simulation set-up}

\label{setup} 

In previous dissipative particle models, \emph{a priori} length and time
scales were fixed by the cut-off radius \cite{hoog92,espanol95,avalos97}.
For this model, the only limiting condition is that the number of molecules $%
N_{k}$ in each DP must be large enough such that $1/{N_{k}}\ll 1$. For the
purpose of the present paper we have chosen to work with an average number
of molecules per dissipative particle larger than 500 hundred. There are of
course no upper bounds to the number of molecules we may assign inside a DP,
but when this number is sufficiently large, the fluctuations disappear and
the model becomes a standard Lagrangian hydrodynamics code. The input
parameters are the size of the simulation box $L$, the number of dissipative
particles $N_{DP}$, the molar mass $mol$, the temperature $T$ and the
density $\rho $. 

In our three-dimensional simulations, we set up the parameters appropriate
for argon. A set of numerical values for water and argon are expressed in
cgs units in Table \ref{tableinput}. The equation of state contains the
contribution of the intermolecular potential in the pairwise approximation.
The size of the simulation box is set up depending on the number of DPs and
the number of molecules per DP and the tessellation initialised with the
Voronoi tessellation composed of regular hexagons or a random configurations
of points (three dimensions) in periodic boundary conditions. %

We ran a set of equilibrium simulations to benchmark the code in order to
establish the computational limits of the model and the Voronoi tessellation
on our machine. Because of the thermodynamic consistency of the model, the
temperature obtained from the internal energy of the fluid particles is
equal to the kinetic temperature, i.e. the temperature of the set of
particles, computing the total kinetic energy with zero mean and converting
the internal energy to the temperature. We use the set of equations (\ref%
{sdeiso}). In three dimensions, the equilibrium simulations are performed
starting from an isothermal-isobaric configuration with fixed internal
energy corresponding to $300K$ for argon but with zero momentum and equal
pressures for all fluid particles. The number of fluid particles is 1000,
while the number of molecules per fluid particle is set by varying the size
of the system. In these simulations, we set an average of about 500
molecules per DP and a three dimensional simulation box with side $2650
\times 10^{-8}cm$. The simulations showed that the DPD temperature of the
system increases until it reaches 300K beyond which value the dissipative
term causes the extra kinetic energy to be dissipated, operating like a
thermostat. This gives strong support for the fluctuation-dissipation
relations in three dimensions as well, confirming for the three-dimensional
model what has been tested already in two dimensions \cite%
{flekkoy00,serrano01}. The temperature is one of the first quantities to
reach equilibrium. For momentum, mass density, pressure, volume, etc. the
equilibrium distributions can be analytically computed and the predictions
confirmed via a more extensive series of numerical simulations which have
been reported elsewhere \cite{serrano02}.

\section{Parallel implementation and performance in two dimensions}

Whereas a parallel code can be written in two and three spatial dimensions
following exactly the same strategy, we have so far implemented and tested
the parallel performances of the model only in two dimensions. In fact, the
parallel code uses the same Algorithm 1 to isolate boundary points as the
sequential code. However, whereas in the sequential code the image points
are the points in the boundary region of the opposite face, in the case of
the parallel implementation they are the boundary points on an adjacent
processor. The use of image points gives a straightforward parallel
implementation with a message passing paradigm that in this case has been
accomplished using the MPI (Message Passing Interface) \cite{mpi} standard
library. The big advantage of the image points solution is that this
parallelisation is independent of the library used. The library is used as a
software engine to compute the tessellation only.

The model contains three phases of communication:

\begin{itemize}
\item to construct the tessellation using the sequential code for periodic
boundary conditions, but communicating the ghost region to the adjacent
processor;

\item to move the particles to the corresponding processor when the
particles fall outside the processor's domain due to the position update;

\item to preserve the force symmetry between two particles in different
domains due to the stochastic force term in the Langevin equations.
\end{itemize}

\begin{figure}[!tb]
\vspace*{.05in} \centerline{\mbox{\epsfig{file=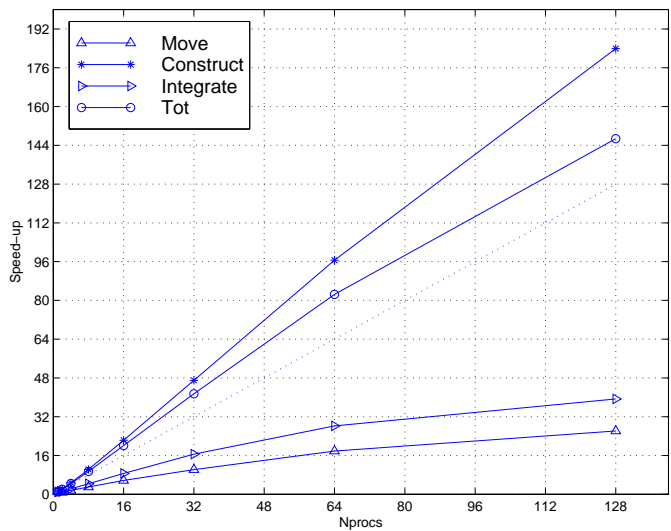,height=7cm}}}
\caption{ Speed-up index on a Cray T3E. The dotted line indicates the linear
speed-up. Circles indicate overall speed-up, stars Voronoi construction
speed-up, triangles the speed-up of the integration and update of the
Langevin equations. }
\label{spt3e}
\end{figure}

The performance analysis of the code is reported for a Cray T3E (256
processors Alpha 600MHz) in Figure \ref{spt3e} and Table \ref{timet3e}. All
the benchmarks are run with the same configuration of $128000$ fluid
particles (Npart) for $50$ iterations.

%

The simulation is a relaxation towards equilibrium of the isothermal model
\cite{flekkoy00} starting from a random initial configuration. The timings
have been taken measuring the CPU-clock time for the routines: \textit{move}
updates the position and communicates the particle outside the boundary,
\textit{construct} constructs the tessellation, \textit{integration}
integrates the Langevin equations and communicates the stochastic force,
\textit{comm} is the total time of communication without the time for
buffering and \textit{total} measures the total time without Input-Output.
All the times measured are inclusive of communication, which is reported
separately for comparison; the time for one processor is the time of the
sequential version without any overhead due to buffering.

The important point which we wish to stress here is that although the
construction and maintenance of the Voronoi tessellation represents a
significant computational overhead due to the algorithmic complexity of $%
NlogN$, with this parallel implementation the speed-up obtainable is very
promising. This is an intrinsic feature of the model. Each fluid particle
interacts only with its nearest neighbour, thereby reducing the amount of
communication. Previous standard dissipative particle methods \cite%
{espanol95,monaghan92} lacked this strong locality feature, interacting with
many more neighbouring particles and severely limiting the parallel
performance \cite{jury99,novik00}.

Looking at the speed-up index, it is worth noting that the super-linearity
shown was expected because the algorithmic complexity of the construction is
not linear, but $Nlog(N)$. The construction of the tessellation for $N/2$
takes less than half of the time spent for $N$ dissipative particles.

The present parallel implementation may suffer from load imbalance depending
on the nature of the multiscaling in a given application. If the length
scale is set by the polymer molecule and the polymer density is uniform in
the simulation region then the parallel implementation should be well
balanced. A different situation may arise in the case of multiscaling due to
the flow structure, when some processors may have a higher computational
load than others. In this case the parallel performance will drop to the
slowest processor's performance. A dynamic repartitioning of the simulation
region among processors should reduce this effect, but the repartitioning
routine would necessitate further communication, thereby producing an
additional overhead.

\section{Computational limits}

In the preceding sections, we described several algorithms that we have used
to implement boundary conditions and dynamic reconstruction of the
tessellation. We showed as well the good parallel performance of the model
due to the locality of its interactions. Now we want to estimate the
computational limits of this model in view of future applications to complex
fluids flows. The CGAL library version 2.3 is used for these runs.

In two dimensions, a sequential code can easily handle 50,000 fluid
particles. Considering the good parallel performance, it is possible to run
very large scale simulations. The situation in three dimensions is quite
different. The poor efficiency of periodic boundary conditions limits the
overall performance of the code. In Table \ref{tableperf}, we report the
computational time in seconds for a set of runs with increasing number of
fluid particles, for 10 iterations. The simulation is a simple equilibration
of the fluid particle system. We also report the average number of
topological events (reconnections) that are performed for each iteration,
the time to move and maintain all the fluid particles, the Eulerian time
corresponding to the time needed to solve the Langevin equations once the
tessellation and the volumes have been computed, and finally the time to
impose boundary conditions.

%
Table \ref{tableperf} suggests that while the maintenance of the
tessellation is reasonable compared to the time needed to integrate the
Langevin equations, the boundary conditions do not allow to increase the
size of the simulation to more than few thousands fluid particles. For the
maintenance task, a non-local reconnection algorithm, which restores the
Delaunay property after the entire set of points are moved, would be at
least 3 times faster in the move task. However, the stability would be
affected and the possibility of asynchronous time update is also lost.
Instead, our code proved to be very stable. Further optimisations are of
course possible using local flips and remove-insert procedures only in
pathological configurations.

\section{Conclusions}

Within the framework of fluctuating hydrodynamics codes for complex fluids,
we have described our present algorithmic approach to multiscale dissipative
particle dynamics. The algorithms to dynamically maintain the Voronoi
tessellation, impose boundary conditions and develop parallel codes for the
Voronoi dissipative particle dynamics model have been described in two and
three spatial dimensions. The use of publicly available  computational
libraries \cite{triangle,cgal} is an advantage compared to direct coding
because the evolution of the library is automatically incorporated into the
model. This is an important factor at this stage of research in
computational geometry where a really stable and efficient library is still
under development. In this context, CGAL offers a first interesting
approach, delivering a usable and powerful tool for scientists.

The reconnection algorithm used to dynamically maintain the tessellation in
two dimensions is stable as long as it is applied on a valid albeit not
Delaunay triangulation. This is not easy to guarantee even reducing the time
step. In three dimensions, our approach in Section \ref{rec3d} has a limited
additional cost, but offers the advantage of great stability. The overall
efficiency is good for running simulations in real applications.

Boundary conditions are evidently not yet efficiently addressed in three
dimensions. Periodic boundaries take most of the time in the simulation. The
image points approach for boundaries is appealing because it provides at the
same time periodic, fixed wall boundary and a domain decomposition parallel
implementation, but it does not perform efficiently due to the large number
of image points and the complexity and cost of the point removal from the
tessellation in three dimensions. However, it is possible to use periodic
boundary in the $x$ direction and a wall surface boundary on the $y$ and $z$
to significantly reduce the amount of image points. This corresponds to
flowing a fluid in a duct, where shear flows can be applied to compute the
rheological properties of the fluid.

Alternative implementations of three dimensional periodic boundary
conditions can be envisaged. Although, we have not actually implemented it,
it is nevertheless interesting to consider here an alternative option.
Imposing periodic boundary conditions on a bounded region of space can be
thought as actually bending the Euclidian space of the region to form a
torus. This is actually equivalent to changing the topology of the region
which, in fact, might lead to problems of finite size effects (in case the
triangulation becomes locally non-euclidian). We have implemented boundary
conditions by imposing periodicity on the Euclidian space around a box of
fixed size. This makes use of image points outside the box in order to
obtain the periodic boundary. A second alternative is to use a three
dimensional torus, without image points, but directly connecting the fluid
particles from one side to the other. Of course, this necessitates some care
when handling the coordinates. A possible implementation of this solution
requires changing the insphere function which is used to construct the
tessellation. This is related to the  distance function, which could be
implemented as $ d_{torus}(\mathbf{P}_1,\mathbf{P}_2):=norm((\mathbf{P}_1-%
\mathbf{P}_2) mod ({\ \mathbf{L}}/2)), $ where $\mathbf{L}=\{L_x,L_y,L_z \}$
represents the three linear dimensions of the box and the modulus operation $%
mod$ is applied to each component. This function must consider the toroidal
topology, in order that points close to the boundary of the previous
Euclidian box are still considered close as they are in the torus. This
implementation would require substantial modifications of the CGAL library.

\begin{ack}
G.D.F. thanks Queen Mary,  University of London,  and  Schlumberger Cambridge
Research for funding his Ph.D studentship. We thank CGAL for
help and support, and  P. Espa{\~n}ol and M. Serrano for useful discussions
and exchange of preprints.
\end{ack}

\section*{A Maintaining the Voronoi tessellation}

\label{codeapp}

In two dimensions, we are able to restore the Delaunay triangulation after
every time-step with the algorithm listed below, which is an extension of
the restore\_delaunay function in Delaunay\_triangulation\_2.h. We first
compute the time-step such that no Delaunay vertex can move outside its
Voronoi region. This helps to ensure that we have a valid triangulation, but
it does not guarantee it. We integrate the stochastic Langevin equations of
motion and then update the position of each vertex by calling the CGAL
set\_point function. For each vertex \textit{v}, we then circulate around
the incident faces, checking that \textit{v} is not located inside the
circumcircle of each neighbour of \textit{v} (external flip) and that no
vertices adjacent to \textit{v} lie inside the circumcircle of each face
(internal flip). For any invalid Delaunay triangle, we simply triangulate
the quadrilateral the other way by calling the CGAL flip function. Thus we
are able to efficiently restore the two dimensional Voronoi tessellation
locally.
\begin{verbatim}

void restore_delaunay(Vertex_handle vh) {
        int i;
        Face_handle f=vh->face(),next,start(f);
        do {
                i=f->index(vh);
                if(!is_infinite(f)) {
                   if(!internal_flip(f,cw(i))) external_flip(f,i);
                   if(f->neighbor(i)==start) start=f;
                }
                f=f->neighbor(cw(i));
           } while(f!=start);
}


void external_flip(Face_handle& f, int i) {
        Face_handle n=f->neighbor(i);
        if(ON_POSITIVE_SIDE !=
           side_of_oriented_circle(n,f->vertex(i)->point())) return;
        flip(f,i);
        external_flip(f,i);
        i=n->index(f->vertex(i));
        external_flip(n,i);
}


bool internal_flip(Face_handle& f, int i) {
        Face_handle n=f->neighbor(i);
        if(ON_POSITIVE_SIDE !=
           side_of_oriented_circle(n,f->vertex(i)->point())) return false;
        flip(f,i);
        return true;
}

\end{verbatim}

In three dimensions, the in\_sphere property is checked for all the facets
incidents to a given vertex vh. If a topological event is found then the
point is removed and inserted again. The facets are inserted in a set in
order to avoid double checking the same facet. An implementation is listed
below.
\begin{verbatim}

typedef std::set<Facet,compare_facets> Facets_cont;

bool move_reconnection(Vertex_handle& vh,const Vector& dr) {
    Point old_point=vh->point();
    Point new_point=old_point;
    move(new_point,dr);
    vh->set_point(new_point);
    if (check_topological_event(vh))
        {vh->set_point(old_point);move(vh,dr);return false;}
    return true;
    }

bool check_topological_event(const Vertex_handle& vh) const {
    Facets_cont facets;
    composing_facets(vh,facets);
    Facets_cont::iterator fi=facets.begin(),fiend=facets.end();
    for (;fi!=fiend;fi++) if (facet_in_sphere(*fi)) return true;
    return false;
}


void composing_facets(const Vertex_handle& vh,Facets_cont& facets)
const {
    std::set<Cell_handle> cells;
    this->incident_cells(vh,cells);
    std::set<Cell_handle>::iterator ci=cells.begin(),cend=cells.end();
    for(;ci!=cend;ci++)
    {
    facets.insert(Facet(*ci,0));
    facets.insert(Facet(*ci,1));
    facets.insert(Facet(*ci,2));
    facets.insert(Facet(*ci,3));
    }
}

bool facet_in_sphere(const Facet& f) const {
 return (ON_BOUNDED_SIDE==
        side_of_sphere(f.first,
                      (*(f.first->mirror_vertex(f.second))).point()));
}

\end{verbatim}

\bibliographystyle{AIP}
\bibliography{../gianni}

\newpage
\begin{table}[!t]
\caption{ Number of image points inserted in two and three spatial
dimensions to impose boundary conditions.}
\label{imagepoints}\vspace{0.1in}
\centerline{
\begin{tabular}{c c c c c}
\hline
No. of DPs & 2D  images & 3D shell & 3D  images & 3D extra images
\cr
 \hline
 100& 64 & 100 &  408 & 195 \cr
1000 &  190 & 687 & 1329 & 398\cr
10000 &  581 & 3855 & 5287 & 661 \cr
 50000 & 2081 & 11998 & 14370 & 1050\cr
\hline
\end{tabular}
}
\end{table}

\begin{table}[!b]
\caption{ The input parameters for argon and water used for the simulations.
All the numerical values are expressed in cgs units for three spatial
dimensions. $n=N/N_{A}$, $N_{A}$ is Avogadro's number and $E_{id}$ is the
internal energy of an ideal gas.}
\label{tableinput}
\vspace{0.1in}
\centerline{
\begin{tabular}{ l c c }
\hline cgs units & argon & water \\ \hline
temperature & 300 & 300 \\
viscosity & 2.26E-4 & 0.01 \\
heat diffusivity & 2E3 & 5.9E4 \\
mass density & 0.00178 & 1 \\
molecular mass & 39.94 & 18\\
pressure equation of state & $PV=Nk_{B}T$ & $\left(
P+\frac{n^2a}{V^2} \right) \left( V-nb \right)= Nk_{B}T $\\
\hline
\end{tabular}
}
\end{table}
\clearpage

\begin{table}[!t]
\caption{ Timings in seconds of the main routines on a Cray T3E. Comm
measures the total time for communication, while the other timings are
inclusive of the communication time. The number of iterations is 50.}
\label{timet3e}\vspace{0.1in}
\centerline{
\begin{tabular}{c c c c c c  c}
\hline
Nprocs & Npart/NProcs & Move & Construct & Integration & Comm & Total \\
\hline
1 & 128000 & 4.50 & 395.17 & 19.38 & 0 & 427.73 \\
2& 64000 & 5.45 & 180.96 & 16.54 & 1.49 & 206.99 \\
4 &32000 & 2.76 & 83.42 & 8.46 & 1.43 & 96.52 \\
8 & 16000 &1.48 & 38.48 & 4.43 & 1.02 & 45.34 \\
16 & 8000 & 0.78 &17.70 & 2.24 & 0.67 & 21.24 \\
32 & 4000 & 0.44 & 8.40 &1.16 & 0.44 & 10.31 \\
64 & 2000 & 0.25 & 4.09 & 0.68 &0.38 & 5.18 \\
128 & 1000 & 0.17 & 2.14 & 0.49 & 0.47 &2.91 \\
\hline
\end{tabular}
}
\end{table}

\begin{table}[b!]
\caption{ Computational time in seconds for 10 iterations of the three
dimensional tessellations with periodic boundary conditions.}
\label{tableperf}\vspace{0.1in}
\centerline{
\begin{tabular}{ l c c c c}
\hline
Fluid particles & Reconnected & Move & Eulerian & Update images\\
\hline
100 & 5 &0.86 s & 0.87 s& 17.5 s\\
1000 & 70 & 9.5 s & 1.0 s & 65.6 s \\
5000 & 450& 49.0 s & 5.0 s & 201.5 s \\
\hline
\end{tabular}
}
\end{table}

\clearpage

\begin{figure}[tb]
\vspace*{.05in} \centerline{\mbox{\epsfig{file=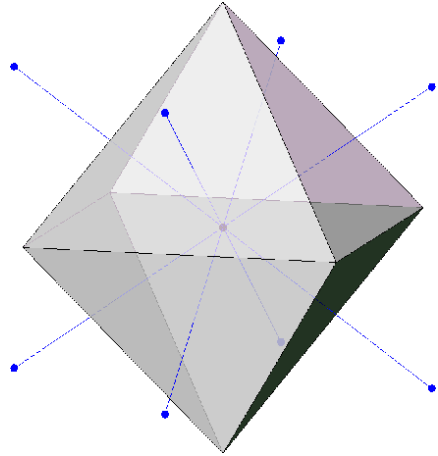,height=6cm}}}
\caption{ A fluid particle in three dimensions for the simple case of a
regular grid. The stresses and fluxes are computed over the surfaces between
the Voronoi centre points. }
\label{fpfigure}
\end{figure}

\end{document}